\documentclass[10pt,conference]{IEEEtran}
\usepackage{algorithm}
\usepackage{algpseudocode}
\usepackage{cite}
\usepackage{amsmath,amssymb,amsfonts}
\usepackage{graphicx}
\usepackage{tikz}
\usetikzlibrary{matrix}
\usetikzlibrary{positioning}
\usepackage{textcomp}
\usepackage{xcolor}
\usepackage{amsthm}
\usepackage{pgfplots}
\pgfplotsset{compat=1.3}
\usepackage{makecell}
\usetikzlibrary{external}
\usepackage{booktabs}
\usepackage{xcolor}

\def\BibTeX{{\rm B\kern-.05em{\sc i\kern-.025em b}\kern-.08em
    T\kern-.1667em\lower.7ex\hbox{E}\kern-.125emX}}

\def\be{\boldsymbol{e}}
\def\bE{\boldsymbol{E}}
\def\bW{\boldsymbol{W}}
\def\dd{\mathrm{d}}
\newtheorem{lem}{Lemma}

\DeclareMathAlphabet\mathbfcal{OMS}{cmsy}{b}{n}
\DeclareMathOperator{\EX}{\mathbb{E}}% expected value

\newcommand{\bSigma}{\boldsymbol{\Sigma}}   
\newcommand{\by}{\boldsymbol{y}}

\newcommand{\bOmega}{\boldsymbol{\Omega}} 
\newcommand{\bomega}{\boldsymbol{\omega}}

\newcommand{\bx}{\boldsymbol{x}} 
\newcommand{\bF}{\boldsymbol{F}}

\newcommand{\bz}{\boldsymbol{z}}

\newcommand{\bX}{\boldsymbol{X}} 
\newcommand{\bD}{\boldsymbol{D}}
\newcommand{\bB}{\boldsymbol{B}}

\newcommand{\bK}{\boldsymbol{K}}

\newcommand{\bA}{\boldsymbol{A}} 
\newcommand{\bI}{\boldsymbol{I}} 
\newcommand{\bY}{\boldsymbol{Y}}

\newcommand{\herm}{^{\mathsf{H}}}

\newcommand{\inv}{^{^{-1}}}

\newcommand{\ex}[1]{\EX\left[{#1}\right]}

\newcommand{\size}[3]{\in \mathbb{{#1}}^{{#2} \times {#3}}}

\newcommand{\tr}[1]{\textrm{tr} \left( {#1} \right)}

\DeclareMathOperator*{\minimize}{\mathrm{minimize}}

\DeclareMathOperator*{\subjto}{\mathrm{subject \, to}}

\newtheorem{prop}{Proposition}

\begin{document}

\title{Fronthaul-Constrained Distributed Radar Sensing}
\author{Christian Eckrich$^{*,**}$, Abdelhak M. Zoubir$^{*}$, and Vahid Jamali$^{**}$\\
$^{*}$Signal Processing Group, $^{**}$Resilient Communication Systems Group\\
Technical University of Darmstadt, Darmstadt, Germany
\thanks{This work has been performed in the context of the LOEWE center emergenCITY [LOEWE/1/12/519/03/05.001(0016)/72].}
\vspace{-0.5cm}}

\maketitle

\begin{abstract}
In this paper, we study a network of distributed radar sensors that collaboratively perform sensing tasks by transmitting their quantized radar signals over capacity-constrained fronthaul links to a central unit for joint processing. We consider per-antenna and per-radar vector quantization and fronthaul links with dedicated resources as well as shared resources based on time-division multiple access. For this setting, we formulate a joint optimization problem for fronthaul compression and time allocation that minimizes the Cramer Rao bound of the aggregated radar signals at the central unit. Since the problem does not admit a standard form that can be solved by existing commercial numerical solvers, we propose refomulations that enable us to develop an efficient suboptimal algorithm based on semidefinite programming and alternating convex optimization. Moreover, we analyze the convergence and complexity of the proposed algorithm. Simulation results confirm that a significant performance gain can be achieved by distributed sensing, particularly in practical scenarios where one radar may not have a sufficient view of all the scene. Furthermore, the simulation results suggest that joint fronthaul compression and time allocation are crucial for efficient exploitation of the limited fronthaul capacity.
\end{abstract}

\section{Introduction}

The co-design of sensing, communication, and computing is one of the key features of the sixth-generation (6G) mobile communication systems \cite{feng2021joint}. An example of this convergence is distributed sensing assisted by cloud/edge processing, where sensing data from various distributed sensors are communicated to a central unit (CU) at the edge/cloud for joint processing \cite{wen2024integrated}. This network architecture benefits from the diversity offered by observations at different sensors as well as the computational power available at the edge/cloud. Distributed sensing is expected to find various applications in emerging technologies such as digital twins, virtual reality, and autonomous driving~\cite{lu2024integrated}. 

Distributed sensing using radars has been investigated in \cite{nanzer2021distributed,Godrich_2011,Garcia_2014,Xie_2018,Zabolotsky_2018,Chalise_2023,Yi_2020,Yan_2021}. In particular, \cite{Godrich_2011,Garcia_2014,Xie_2018} studied resource allocation among distributed radars, namely power allocation \cite{Godrich_2011}, joint power and bandwidth allocation \cite{Garcia_2014}, and joint power and radar selection \cite{Xie_2018}. A distributed signal processing based on sharing the averaging consensus among radars was developed in \cite{Zabolotsky_2018}. Similarly, a distributed average consensus-based estimation was proposed in \cite{Chalise_2023}
where neighboring radars share their generalized likelihood ratio test (GLRT) values. 
However, the works in \cite{nanzer2021distributed,Godrich_2011,Garcia_2014,Xie_2018,Zabolotsky_2018,Chalise_2023} assumed ideal communication links (neglecting capacity constraints or transmission errors) among radars and/or from radars to the CU. To reduce the communication requirements, a distributed fusion architecture was proposed in \cite{Yi_2020} without explicitly accounting for the communication links.  In \cite{Yan_2021}, a distributed sensor network was studied, where sensors share their local estimates of the time-of-arrival with the CU over a noisy wireless channel.

The 6G communication networks will unlock the full potential of distributed sensing networks by joint processing of the radar signals (not their local estimates) at the CU exploiting edge/cloud computing. In practice, the real-valued radar signals must be quantized so that they can be transmitted to the CU over capacity-constrained fronthaul links. Fronthaul compression has been widely investigated in the context of cloud radio access networks (C-RAN) \cite{zhou2014optimized,peng2015fronthaul,najafi2019c}, where the objective is the joint detection of the data of multiple users at the CU. Due to fundamental differences in the performance measures of detection and sensing problems (e.g., the data rate vs. estimation accuracy), the methods developed in \cite{zhou2014optimized,peng2015fronthaul,najafi2019c} cannot be directly applied to distributed radar sensing.
Fronthaul compression has been recently considered in  \cite{Zhang_2024} for collaborative edge artificial intelligence (AI) inference using over-the-air computation technique, where the discriminant gain was adopted as a performance metric. To the best of the authors' knowledge, the optimization of fronthaul compression for distributed radar networks based on estimation-theoretic performance measures has not been studied in the literature,~yet.
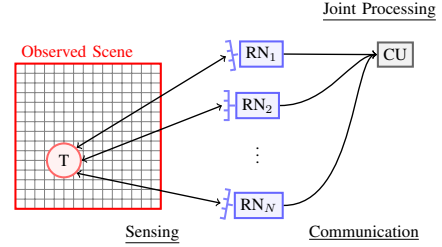
\begin{figure}
    \centering
    \resizebox{0.65\linewidth}{!}{%
    \begin{tikzpicture}[
    targetnode/.style={circle, draw=red!60, fill=red!5, very thick, minimum size=7mm},
    radarnode/.style={rectangle, draw=blue!60, fill=blue!5, very thick, minimum size=5mm},
    fusionnode/.style={rectangle, draw=black!60, fill=black!5, very thick, minimum size=5mm},
    radararray/.style={
        %draw=blue!60,
        %fill=blue!5,
        very thick,
        minimum height=0.7cm,
        path picture={
            \foreach \y in {-0.3,-0.1,0.1,0.3}{
                \draw[blue!50] 
                (0,0) ++(-0.1,\y) -- ++(0.1,0);
            }
           \draw[blue!50] 
                (0,-0.32) -- ++(0,0.64);
            \draw[blue!50] 
                (0,0) -- ++(2,0);
        },
    },
    ]
    %Nodes
    \draw[gray,step=2mm] (-1,2) grid (2,-1);
    \draw[red, very thick] (-1,2) rectangle (2,-1);
    \node[align=left, anchor=south west] at (-1,2) {\textcolor{red}{Observed Scene}};
    \node[targetnode]        (T1)       [] {T};

    \node[radararray]      (R2S)       [above right=0.5cm and 3cm of T1] {};
    \node[radarnode]      (R2)       [right=0cm of R2S] {$\text{RN}_2$};
    \node[radarnode]      (R1)       [above left=0.5cm and -1cm of R2] {$\text{RN}_1$};
    \node[radararray,rotate=10,anchor=east]      (R1S)       [left=0cm of R1] {};
    \node[radarnode]      (R3)       [below left=1.5cm and -1cm of R2] {$\text{RN}_N$};
    \node[radararray,rotate=-10,anchor=east]      (R3S)       [left=0cm of R3] {};
    \node[] (D) [below=0.25cm of R2]{$\vdots$};
    \node[fusionnode]      (FC)       [above right=0.5cm and 2cm of R2] {CU};
    
    \node[] (Sens) [below left= 0cm and 1cm of R3] {\underline{Sensing}};
    \node[] (Com) [below right=0cm and 0.4cm of R3] {\underline{Communication}};
    \node[] (Com) [above right=0.3cm and -2cm of FC] {\underline{Joint Processing}};
    
    %Lines
    \draw[<->,thick] (T1.north east) -- (R1S.west);
    \draw[<->,thick] (T1.east) -- (R2S.west);
    \draw[<->,thick] (T1.south east) -- (R3S.west);
    \draw[->,thick] (R1.east) .. controls +(right:7mm) and +(-180:10mm) .. (FC.west);
    \draw[->,thick] (R2.east) .. controls +(right:10mm) and +(-180:5mm) .. (FC.west);
    \draw[->,thick] (R3.east) .. controls +(right:15mm) and +(-180:5mm) .. (FC.west);
\end{tikzpicture}}
    \caption{Schematic depiction of the observed scene containing the target $\text{T}$, $N$ radar sensor nodes (RNs), and the CU. \vspace{-0.3cm}}
    \label{fig:Setup}
\end{figure}
Focusing on fronthaul-constrained distributed radar networks, this paper makes the following contributions:
\begin{itemize}
\item We consider a network of distributed radar sensors that collaboratively perform sensing tasks by transmitting their quantized radar signals over capacity-constrained fronthaul links to a CU for joint processing as depicted in Fig. \ref{fig:Setup}. We consider per-antenna and per-radar vector quantization and fronthaul links with dedicated resources as well as shared resources based on time-division multiple access (TDMA). As far as the authors are aware, this setting has not been studied in the literature before.
\item We formulate a joint optimization problem for fronthaul compression and time allocation that minimizes the weighted Cramer Rao bound of the aggregated radar signals at the CU. The weights are design parameters that can be chosen to prioritize the estimation accuracy of a desired subarea in the overall scene.
\item Since the formulated optimization problem does not admit a standard form that can be solved by existing commercial numerical solvers, we propose refomulations that enable us to develop an efficient algorithm based on semidefinite programming (SDP) and alternating convex optimization (ACO). 
\item We analyze the convergence behavior of the proposed algorithm and its computational complexity as a function of system parameters, i.e., numbers of radars and antennas. 
\item Finally, we present various simulation results to evaluate the performance of the proposed algorithm. These results suggest that a significant performance gain can be achieved by distributed sensing, particularly in practical scenarios where one radar may not have a sufficient view of the entire scene. Furthermore, joint fronthaul compression and time allocation become particularly crucial when the fronthaul capacity is more limited.
\end{itemize}

\textit{Notation:} We adopt the following notation throughout the paper. Bold lowercase and uppercase letters represent vectors and matrices, respectively. $\bI$ denotes the identity matrix and $\bA\succeq 0$ implies that $\bA$ is a positive semidefinite matrix. Moreover, $|\cdot |$, $\tr{\cdot}$, $[\cdot]\herm$, $[\cdot]\inv$, $[\cdot]^+$, and $\textrm{vec}[\cdot]$ denote the determinant, trace, hermitian transpose, inverse,  generalized inverse, and column-wise verctorisation of a matrix,  respectively. $\mathbb{R}$ and $\mathbb{C}$ are the sets of complex and real numbers, respectively. $\mathcal{CN}\left(\boldsymbol{\mu},\bSigma\right)$ represents a multivariate complex Gaussian random variable (RV) with mean vector $\boldsymbol{\mu}$ and covariance matrix $\bSigma$.  
Finally, $\ex{\cdot}$ denotes expectation and $\mathcal{O}(\cdot)$ represents the Big-O notation.

\section{Signal Model}

In this section, we introduce the  models adopted in this paper for radar sensing signals, fronthaul compression, fronthaul capacity, and the overall sensing measure at the CU.

\subsection{Radar Signal Model}
We consider full-duplex monostatic radars equipped with one antenna to send the radar signal and $M$  antennas to receive the reflections. The total bandwidth allocated to the sensing signals is denoted by $B_R$ and is equally shared among the $N$ radars to avoid any crosstalk. Furthermore, we assume each radar employs a stepped frequency waveform spanning $K$ frequencies. Thereby, the sensing frequencies adopted by the $n$th radar is given by $f_{n,k}=f_n+(k-1)B_R/NK,\,\,k=1,\dots,K$, where $f_n=f_0+(n-1)B_R/N$ is the initial frequency used by the $n$th radar and $f_0$ is the beginning of frequency band used for sensing signals.
The entire scene of interest is divided into $G$ equally spaced grid cells. 
Let $\tau_{n,m,g}$ denote the time it takes for the electromagnetic wave to travel from the transmit antenna to the grid cell $g$ and back to the receive antenna $m$ of radar $n$. The received signal at antenna $m$ of radar $n$ for step frequency $k$ is modeled as
\begin{align}
    y_{n,m,k} = \sum_{g=1}^{G} \sqrt{P_n} h_{n,m,g}   e^{-j 2 \pi f_{n,k} \tau_{n,m,g}} x_g +\omega_{n,m,k},
\end{align}
where $P_n$ denotes the transmit power of radar $n$, $h_{n,m,g}$ represents the end-to-end channel free-space channel attenuation, i.e., $h_{n,m,g}\propto(\tau_{n,m,g}\cdot c)^{-2}$ with $c$ being the speed of light, and  $x_g$ models the target radar cross section (RCS) with expected power $\ex{|x_g|^2}=\sigma^2$. Moreover, $\omega_{n,m,k}$ represents the zero-mean additive white Gaussian noise (AWGN) at the $n$th radar, i.e., $\mathcal{CN}(0,\sigma_n^2)$, where $\sigma_n^2$ denotes the receive noise power at antennas of the $n$th radar.

For compact representation, the captured signals at all receive antennas of radar $n$ and all $K$ step frequencies are collected in the signal matrix $\bY_n \size{C}{K}{M}$, which can then be vectorized to form a long measurement vector $\by_n = \textrm{vec}\left[\bY_n\right]\size{C}{KM}{1}$ given by
\begin{align}
    \by_n &= \bA_n \bx +\bomega_n.\label{Eq:VectorNotation}
\end{align}
Here, $\bA_n\size{C}{KM}{G}$ is called the measurement matrix of the radar $n$, where the element on the $i$th row (corresponding to the $k$th step frequency and $m$th antenna) and $g$th column is given~by 
\begin{align}
    [\bA_n]_{i,g} = \sqrt{P_n}h_{n,m,g}e^{-j 2 \pi f_{n,k} \tau_{n,m,g}}.
\end{align}
 Moreover, in \eqref{Eq:VectorNotation}, $\bx\size{R}{G}{1}$ collects the RCS of the target at all $G$ grid points. Furthermore, $\bomega_n\size{C}{KM}{1}$ collects all observation noises at the $n$th radar, where $\bomega_n \sim \mathcal{CN}\left(\boldsymbol{0},\bOmega_n\right)$, with $\bOmega_n=\sigma_n^2 \bI$.
 
\subsection{Fronthaul Compression}
The radars transmit a sampled and quantized version of their received signal $\by_n$ to the CU.   Using rate-distortion theory, the quantized signal, denoted $\hat{\by}_n$ can be modeled by the Gaussian test channel  as follows \cite{cover_1991}
\begin{align}
    \hat{\by}_n = \by_n +\bz_n 
    = \bA_n \bx +\bomega_n +\bz_n,
\end{align}
where the distortion noise $\bz_n=\hat{\by}_n-\by_n\size{C}{KM}{1}$ is distributed as $\mathcal{CN} \left(\boldsymbol{0},\bD_n\right)$, where $\bD_n = \ex{\bz_n\bz_n\herm}$ represent the distortion covariance matrix. Let $r_n$ denote the output rate of the quantizer at radar $n$. The rate-distortion theory establishes a lower bound on  $r_n$. We present this lower bound for two cases depending on whether or not the correlation among signals at each radar is exploited. 

\textit{1. Per Antenna/Frequency Vector Quantization (AFVQ):}
In this case, the signals from each radar antenna and step frequency are quantized separately. The total  rate of the quantizers at the $n$th radar must meet
\begin{align}
r_n\geq &\sum_{m=1}^M\sum_{k=1}^K f_s \log  \left( \frac{\sum_{g=1}^GP_n|h_{n,m,g}|\sigma^2+\sigma_n^2+d_{n,m,k}}{d_{n,m,k}} \right)\nonumber \\
&= f_s \log \left( \frac{\left|\bK_n  +\bD_n\right|}{\left|\bD_n\right|} \right)\triangleq R_n(\bD_n),
    \label{eq:AVQRate}
\end{align}
where $f_s$ is the frequency that radar signal is sampled, and $d_{n,m,k}$ is the quantization noise variance for the signal received at the $k$th step frequency, $m$th antenna of  $n$th radar. Here, for AFVQ, $\bK_n$  and $\bD_n$ are diagonal matrices with entries $\sum_{g=1}^GP_n|h_{n,m,g}|\sigma^2+\sigma_n^2$  and  $d_{n,m,k},\,\,\forall m,k$, respectively.

\textit{2. Per Radar Vector Quantization (RVQ):}
In this case, each radar quantizes the signals from their antennas and different step frequencies in a joint manner. This leads to efficient exploitation of the correlation that exists among the received signals at each radar. Under RVQ, the output rate of the quantizer at $n$th radar is lower bounded as
\begin{align}
    r_n  &\geq f_s \log \left( \frac{\left|\bA_n \bA_n\herm \sigma^2 + \bOmega_n +\bD_n\right|}{\left|\bD_n\right|} \right)\triangleq R_n(\bD_n).
    \label{eq:RVQRate}
\end{align}
Defining $\bK_n=\bA_n \bA_n\herm \sigma^2+ \bOmega_n$ in the above equation, $R_n(\bD_n)$ assumes the same form for both RVQ and AFVQ except that  $\bK_n$ and $\bD_n$ are diagonal for AFVQ but they are not necessarily diagonal for RVQ.

\subsection{Fronthaul Link Capacity}

We consider the following two cases for fronthaul links:

\textit{1. Dedicated Fronthaul:}
In this case, each radar node is connected to the CU by a dedicated fronthaul link (e.g., an optical fiber) with capacity $C_n,\,\,\forall n$. For the quantized signals to be reliably transmitted over the fronthaul link, the output rate of the quantizer $r_n$ must meet:
\begin{align}
    r_n \leq C_n, \quad \forall n = 1,\hdots,N.
\end{align}

\textit{2. Shared Fronthaul:}
The fronthaul links may share the same time/bandwidth resource, e.g., in wireless fronthauling. Here, we adopt a TDMA protocol, where the $n$th radar employs the fronthaul bandwidth for $a_n\in[0,1]$ fraction of the time, where $\sum_{n=1}^N a_n=1$ must hold.  This leads to the following constraint on the quantizer output rate $r_n$:
\begin{align}\label{Eq:SharedCapacity}
    r_n \leq a_n C_n, \quad \forall n = 1,\hdots,N,
\end{align}
where $C_n$ is the maximum capacity of the $n$th fronthaul link.

\subsection{Sensing Metric at the CU}

As a sensing design criterion, we adopt the Cramer Rao lower bound (CRLB), which establishes a lower bound on the error variance of any unbiased estimator. Let $\be=\hat{\bx}-\bx\size{R}{G}{1}$ be error vector of estimator $\bx$. For unbiased estimators, i.e., $\ex{\hat{\bx}}=\bx$, the error  covariance matrix $\bE=\ex{\be\be\herm}$ must meet the following condition \cite{poor2013introduction}
\begin{align}
    \bE - \bF^{-1} \succeq 0,\label{Eq:CR}
\end{align}
where $\bF$ is the Fischer information matrix (FIM) of  all observations collected at the CU $(\hat{\by}_1,\dots,\hat{\by}_N)$ and is defined~as
\begin{align}
    \bF &= - \ex{\frac{\dd^2\log\left(f\left(\hat{\by}_1,\dots,\hat{\by}_N|\bx\right)\right)}{\dd^2 \bx}},
\end{align}
where $f\left(\hat{\by}_1,\dots,\hat{\by}_N|\bx\right)$ is the joint probability density function (PDF) of the observations conditioned on $\bx$. Using \eqref{Eq:CR}, the CRLB established the following lower bound on the mean square error (MSE) $\ex{\|\be\|^2}$ of unbiased estimators:
\begin{align}
\ex{\|\be\|^2} = \tr{\bE} \geq  \tr{\bF^{-1}} \triangleq \text{CRLB}. \label{Eq:CRLB}   
\end{align}

We consider an extension of the above CRLB expression as the sensing design metric, which allows us to prioritize the sensing accuracy at a desired area within the overall scene. Let $w_g\in[0,1]$ denote the normalized weight assigned to grid point $g$, where $\sum_{g=1}^G w_g=1$ holds. The proposed weighted CRLB (wCRLB) is given by
\begin{align}
\sum_{g=1}^G \! w_g \ex{\|[\be]_g\|^2} \!\!=\!\tr{\bW\bE} \!\geq\!  \tr{\bW\bF^{-1}} \!\triangleq\! \text{wCRLB},   \label{Eq:wCRLB}  
\end{align}
where $\bW$ is a diagonal matrix with elements $w_g$. One can adopt the weights to focus the optimization of distributed radar networks on any desired subset of the grid points, which is beneficial for, e.g., tracking targets. When the weights are set as $\bW=\frac{1}{G}\bI$, then the wCRLB in in \eqref{Eq:wCRLB} reduces to the CRLB in \eqref{Eq:CRLB}, and all the grid points are treated equally, which is relevant for initial target detection, when no prior information is available.  

\section{Proposed Fronthaul Compression and Time Allocation Scheme}

In this section, we first formulate a problem for joint optimization of fronthaul compression and time allocation. Subsequently, we develop an efficient algorithm for solving this problem and discuss its convergence and complexity.

\subsection{Problem Formulation}

We formulate the optimization problem for the fronthaul links that use shared resources. This problem and the proposed solution can be also applied to fronthaul with dedicated capacities after some minor modifications (i.e., setting $a_n=1,\,\,\forall n$). In particular, we aim to optimize the radar noise covariance matrices $\bD_n,\,\,\forall n$, and the fronthaul time allocation $a_n,\,\,\forall n$, such that the wCRLB is minimized, i.e.,
\begin{align}\label{Eq:P1}
    \mathbf{P1:} \,\minimize_{\bD_n, a_n,\,\, \forall n} \quad& \text{wCRLB}\\
    \subjto \quad&  \nonumber\\
    \textrm{C1}: \quad &R_n(\bD_n) \leq a_n C_n, \quad \forall n, \nonumber\\
    \textrm{C2}:\quad&\bD_n \succeq 0,\quad \forall n, \nonumber\\
    \textrm{C3}:\quad&\sum_{n=1}^N a_n \leq 1, \quad a_n \geq 0, \quad \forall n. \nonumber
\end{align}
Here, C1 ensures that there exists a valid quantization rate $r_n$ that meets both the source coding constraints in \eqref{eq:AVQRate} (or \eqref{eq:RVQRate}) and the channel coding constraint in \eqref{Eq:SharedCapacity}. Moreover, constraints C2 and C3 enforce the feasible sets of $\bD_n$ and $a_n,\,\,\forall n$, respectively.

The following lemma provides the wCRLB expression in terms of the optimization variables.

\begin{lem}
    The wCRLB for the considered distributed sensing network is given by
    \begin{align}
        \label{Eq:wCRLBexpression}
         \text{wCRLB} = \text{\normalfont tr}\left(\bW(\bA^+( \bD + \bOmega ) (\bA^+)\herm)\right),
    \end{align}
    where $\bD\size{C}{NKM}{NKM}$ and $\bOmega\size{C}{NKM}{NKM}$ are block diagonal matrices with block diagonal entries $\bD_n$ and $\bOmega_n$, $\forall n = 1, \hdots, N$, respectively, and
\begin{align}
    \bA = 
    \begin{bmatrix}
        \bA_1, & \hdots, & \bA_{N} 
    \end{bmatrix}^{T}\size{C}{NKM}{G}.
\end{align}
\end{lem}
\begin{IEEEproof}
Due to the statistical independence of the observations at different radar sensors, the joint FIM can be obtained~as
\begin{align}\label{Eq:jointFIM}
    \bF = \sum_{n=1}^{N} \bF_n = \bA \herm ( \bD + \bOmega ) \inv \bA,
\end{align}
where $\bF_n$ is the FIM of the observations at the radar $n$ and is given by
\begin{align}
    \bF_n \!=\! - \ex{\frac{d^2\log\left(f\left(\by_n|\bx\right)\right)}{d^2 \bx}} \!=\! \bA_n\herm(\bD_n+\bOmega_n)\inv{\bA_n}, 
\end{align}
and  $f\left(\by_n|\bx\right)$ is the  PDF of observation $\by_n$ conditioned on $\bx$, whose logarithm is given by
\begin{align}
        \log(f(\by_n|\bx)) = -\frac{1}{2}\log(2\pi)-\frac{1}{2}\log(\det(\bD_n+\bOmega_n)) \nonumber \\
        \quad -\frac{1}{2}(\by_n-\bA_n\bx)\herm (\bD_n+\bOmega_n)\inv(\by_n-\bA_n\bx).
\end{align}
Taking the inverse of $\bF$ in \eqref{Eq:jointFIM} and substituting in the wCRLB in \eqref{Eq:wCRLB} leads to \eqref{Eq:wCRLBexpression} and completes the proof.
\end{IEEEproof}

The main bottleneck in solving $\mathbf{P1}$ is constraint C1 since it does not assume a standard form that is admissible by standard commercial numerical solvers. In the following, we present a reformulation of $\mathbf{P1}$ that enables us to develop an algorithm that can be implemented using CVX \cite{cvx}.

\subsection{Problem Reformulation}

The following proposition presents an equivalent optimization problem for $\mathbf{P1}$.

\begin{prop}\label{Prop:P2}
    Problem $\mathbf{P1}$ can be equivalently written~as 
    \begin{align}
        \label{eq:P2}
        \mathbf{P2:} \,\minimize_{\bD_n,\bY_n,a_n,\,\, \forall n} \quad & \text{\normalfont tr}\left(\bW(\bA^+( \bD + \bOmega ) (\bA^+)\herm)\right)\\
        \subjto \quad&  \nonumber\\
        \widetilde{\textrm{\normalfont C1}}:\quad& R^{\rm upp}_n(\bD_n,\bY_n) \leq a_n C_n \nonumber \\
        & \textrm{\normalfont C2} \,\, \text{\normalfont and} \,\, \textrm{\normalfont C3} \nonumber\\
        \textrm{\normalfont C4}:\quad& \bY_n \succeq 0,\nonumber 
    \end{align}
    where 
    \begin{align}
        \label{eq:Rupp}
        R^{\rm upp}_n(\bD_n,\bY_n) = &-\log_2\left|\bD_n\right|- \log_2\left|\bY_n\right| \\
        &+\frac{1}{\ln(2)} \text{\normalfont tr}\left(\bY_n \left(\bK_n +\bD_n\right)\right)-\frac{M}{\ln(2)}.\quad\nonumber 
    \end{align}
\end{prop}
\begin{IEEEproof}
Let us first rewrite $R_n(\bD_n)$ as
\begin{align}
    R_n(\bD_n) &= \log\left|\left(\bK_n +\bD_n\right)\right|-\log\left|\bD_n\right|.
    \label{eq:Rate_reform}
\end{align}
Note that $R_n(\bD_n)$ is the difference of two convex functions. We borrow the following lemma from \cite{Li_2013} to rewrite $R_n(\bD_n)$. 
\begin{lem} 
    The following identity holds
    \begin{align}
        \log_2\left|\bX\inv\right| = \max_{\bY\succeq 0} \log_2\left|\bY\right| -\frac{1}{\ln(2)} \text{\normalfont tr}\left(\bY \bX\right)+\frac{M}{\ln(2)}
    \end{align}
    where the optimal solution of the right side is $\bY = \bX\inv$.
    \label{Lem:1}
\end{lem}
Using Lemma \ref{Lem:1}, the rate in \eqref{eq:Rate_reform} can be rewritten as 
\begin{align}
R_n(\bD_n) = & -\log_2\left|\bD_n\right| 
 - \bigg(\max_{\bY\succeq 0} \log_2\left|\bY_n\right| 
\nonumber\\ 
&-\frac{1}{\ln(2)} \tr{\bY_n \left(\bK_n +\bD_n\right)} +\frac{M}{\ln(2)}\bigg).
\end{align}
Therefore, for an arbitrary $\bY_n \succeq 0$, we can establish the following upper bound on $R_n(\bD_n)$:
\begin{align}
    \begin{split}
        R_n(\bD_n)\leq R^{\rm upp}_n(\bD_n,\bY_n),
    \end{split}
    \label{eq:upper}
\end{align}
where $R^{\rm upp}_n(\bD_n,\bY_n)$ is given in \eqref{eq:Rupp}. 

Next, using \eqref{eq:upper}, we prove the equivalence of $\mathbf{P1}$ and $\mathbf{P2}$.  First, from \eqref{eq:upper}, we can conclude that when constraint $\widetilde{\text{C1}}$ holds, constraint C1 must hold, too. In other words, for a given $\bY_n$, $\widetilde{\text{C1}}$ has reduced the feasible set of $\bD_n$ compared to C1. However, when optimizing $\bY_n$ in $\mathbf{P2}$, Lemma \ref{Lem:1} guarantees that there always exists a $\bY_n$ for which $R^{\rm upp}_n(\bD_n,\bY_n)=R_n(\bD_n,\bY_n)$ holds, i.e., $\bY_n = \left(\bK_n +\bD_n\right)\inv$. In other words, the optimal solution of $\mathbf{P1}$ is also contained in the feasible set of $\mathbf{P2}$, which implies their equivalence and completes the proof. 
\end{IEEEproof}

The advantages of the reformulated problem $\mathbf{P2}$ are two-folded: \textit{i)} The problem is bi-convex in $\bY_n$ and $(\bD_n,a_n),\,\,\forall n$. This allows us to use ACO to solve $\mathbf{P2}$. \textit{ii)} The functions in $\mathbf{P2}$ are admissible by commercial solver such as CVX \cite{cvx}.

\subsection{Proposed Algorithm}
Based on $\mathbf{P2}$, we define the following two subproblems.

\textbf{Subproblem 1:} Here, we optimize $(\bD_n,a_n),\,\,\forall n$, for given $\bY_n,\,\,\forall n$. This leads to
\begin{align}
    \label{eq:P2a}
    \textbf{P2a:} \,\minimize_{\bD_n,a_n,\forall n} \quad & \tr{\bW(\bA^+( \bD + \bOmega ) (\bA^+)\herm)}\\
    \subjto \quad&  \widetilde{\textrm{C1}}, \,\textrm{C2} \,\, \text{and} \,\, \textrm{C3}. \nonumber
\end{align}
The above problem is convex and can be solved by CVX \cite{cvx}. Moreover, since $R_n(\bD_n)\leq R^{\rm upp}_n(\bD_n,\bY_n) \leq a_n C_n$ holds for any given $\bY_n \succeq 0$, the solution to \textbf{P2a} is always a feasible solution to the original problem in \textbf{P1}. 

\textbf{Subproblem 2:} Next, we optimize $\bY_n,\,\,\forall n$,  for given  $(\bD_n,a_n),\,\,\forall n$. Since $\bY_n,\,\,\forall n$, do not appear directly in the objective function of \textbf{P2}, finding $\bY_n,\,\,\forall n$,  for given  $(\bD_n,a_n),\,\,\forall n$, is in general a feasibility problem. In other words, $\bY_n,\,\,\forall n$, determine the new feasible set for \textbf{P2a}, hence its best value must yield the largest feasible set. This leads to
\begin{align}
    \label{eq:P2b}
    \textbf{P2b:} \,\minimize_{\bY_n} \,\, & R^{\rm upp}_n(\bD_n,\bY_n)\\
    \subjto \quad& \textrm{C4}. \nonumber
\end{align}
Using Lemma~\ref{Lem:1}, the solution to \textbf{P2b} can be found in closed form as
\vspace{-0.3cm}
\begin{align}
\bY_n = \left(\bK_n +\bD_n\right)\inv.
    \label{eq:Yopt}
\end{align}
Algorithm \ref{alg:TDMA} summarizes the proposed algorithm for fronthaul compression and time allocation. 
\begin{algorithm}[t]
\caption{Fronthaul Compression and Time Allocation.}\label{alg:TDMA}
    \begin{algorithmic}[1]
        \State $i \gets 0$
        \State $\bD_n^{(i)}\gets d_n^{(0)}\bI, \,\, \forall n$, where for $d_n^{(0)}$, \eqref{eq:init} holds with equality.
        \Repeat
            \State $i \gets i+1$
            \State $\bY_n^{(i)} \gets (\bK_n+\bD_n^{(i-1)})^{-1}, \quad \forall n$
            \State $\bD_n^{(i)}, a_n^{(i)} \gets$ Solve $\textbf{P2a}$ given $\bY_n^{(i)}, \quad \forall n$ 
        \Until{convergence of $\bD_n, a_n,\forall n$}\\
        \Return{$\bD_n^{(i)}, a_n^{(i)}, \forall n$}
    \end{algorithmic}
\end{algorithm}

\subsection{Discussion}\label{sec:discussion}

In the following, we discuss the initialization, convergence, and complexity of Algorithm~\ref{alg:TDMA}.

\textit{1. Initialization:} We initialize Algorithm~\ref{alg:TDMA} with a feasible solution. In particular, we adopt AFVQ quantization scheme with identical distortion values across each radar's antennas and frequency steps, i.e., $\bD_n=d_n\bI$, and uniform fronthaul time allocation, i.e., $a_n=\frac{1}{N},\,\,\forall$. The value of $d_n$ is found to ensure that constraint C1 holds, i.e., 
\begin{equation}
    R_n(\bD_n) = f_s\log_2|d^{-1}_n\bK_n +\bI| \leq \frac{C_n}{N}.
    \label{eq:init}
\end{equation}

We use the following lemma to find an analytical expression for $d_n$ that meets \eqref{eq:init}.

\begin{lem} 
    For any matrix $\bB \succeq 0$, the following identity holds
    \begin{align}
        \log|\bB| &\leq \text{\normalfont tr}\left(\bB-\bI\right),
        \label{eq:lem2}
    \end{align}
    where the inequality holds with equality for $\bB = \bI$.
    \label{Lem:2}
\end{lem}
Applying Lemma~\ref{Lem:2} to \eqref{eq:init}, we obtain:

\begin{equation}
    R_n(\bD_n) \leq  f_s \text{\normalfont tr}\left( d^{-1}_n\bK_n\right) \leq \frac{C_n}{N}
    \,\,\rightarrow\,\,
    d_n \geq \frac{Nf_s \text{\normalfont tr}\left( \bK_n\right)}{C_n}.
    \label{eq:init_final}
\end{equation}
In other words, choosing $d_n = \frac{N\tr{K_n}}{C_n}$ leads to a feasible solution. This value is used as a starting point to gradually decrease $d_n$ until \eqref{eq:init} holds with equality. We refer to this solution as $d_n^{(0)}$ and use it to initialize Algorithm~\ref{alg:TDMA}.

\textit{2. Convergence Analysis:} Let us define the value of the objective function in iteration $i$ by $f^{(i)}=\tr{\bW(\bA^+( \bD + \bOmega ) (\bA^+)\herm)}$. The following proposition formally characterizes the convergence of the proposed algorithm. 

\begin{prop} 
    Algorithm \ref{alg:TDMA} produces a series of non-increasing objective values $f^{(i)}$, i.e., $f^{(i+1)}\leq f^{(i)}$.
    \label{Prop:Convergence}
\end{prop}
\begin{IEEEproof}
Let $(\bD_n^{(i)},a_n^{(i)}),\,\,\forall n$, denote the solution to \textbf{P2a} at iteration $i$. By updating $\bY_n^{(i+1)} = (\bK_n+\bD_n^{(i)})\inv$ as the solution to \textbf{P2b}, we obtain  
\begin{align}
    R^{\rm upp}_n(\bD_n^{(i)}, \bY_n^{(i+1)})\leq R^{\rm upp}_n(\bD_n^{(i)}, \bY_n^{(i)}) \leq a_n^{(i)} C_n,
\end{align}
which implies that the solution $(\bD_n^{(i)},a_n^{(i)}),\,\,\forall n$, of \textbf{P2a} in the $i$-th iteration remains in the feasible set of \textbf{P2a} in the $(i+1)$th iteration. Since \textbf{P2a} is a convex problem, its global optimum solution in the $(i+1)$th iteration must yield a cost function value that is at least as small as that in the $i$th iteration, i.e., $f^{(i+1)}\leq f^{(i)}$. This completes the proof.   
\end{IEEEproof}
The non-increasing sequence of objective values generated by Algorithm \ref{alg:TDMA} guarantees its convergence to a stationary point. 
\textit{3. Complexity Analysis:}
For the RVQ scheme, the first subproblem \textbf{P2a} is an SDP problem with $N$ semidefinite cone constraints with semidefinite variables of dimension $KM$. The (worst-case) computational complexity to find an $\epsilon$-optimal solution is given by $\mathcal{O}\big(\big(N(KM)^{3.5}+N^2(KM)^{2.5}+N^3(KM)^{0.5}\big)\log\frac{1}{\epsilon}\big)$\cite{Boshkovska_2018},\cite[Theorem 3.12]{bomze_2010}. 
The second subpoblem \textbf{P2b} can be expressed in a closed form \eqref{eq:Yopt} where the matrix inversion has a complexity of $\mathcal{O}(N(KM)^3)$. 
For the AFVQ scheme, the complexity reduces because the semidefinite constraint C2 simplifies to a linear constraint and the matrix inversion in \eqref{eq:Yopt} reduces to scalar inversion. Nonetheless, the worst-case complexity of solving \textbf{P2a} using, e.g., interior-point methods, still scales as $\mathcal{O}\left(N(KM)^{3.5}\log\frac{1}{\epsilon}\right)$ \cite[Theorem 4.2]{bomze_2010}. 
Furthermore,  the overall complexity of Algorithm~\ref{alg:TDMA} scales linearly with the number of iterations required for convergence. 

\section{Simulation Results}
In this section, we first introduce the simulation setup and subsequently present our simulation results.
\subsection{Simulation Setup}
\begin{figure}[t]
\centering
\begin{tikzpicture}[scale=0.7]
\begin{axis}[enlargelimits=false, axis on top, axis equal image,ylabel={y coordinate (m)},xlabel={x coordinate (m)}]
\addplot graphics [xmin=0,xmax=7,ymin=0,ymax=2] {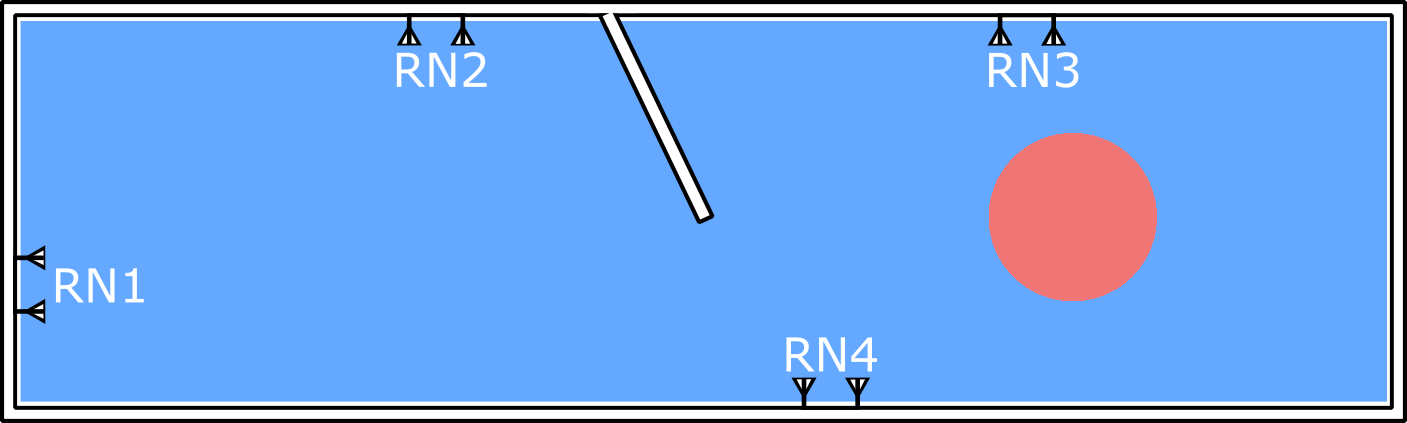};
\end{axis}
\end{tikzpicture}
\caption{The schematic illustration of the locations of the radar nodes in a room setup. The white line depicts a blockage caused by e.g. a wall. We consider two cases for the area of interest, namely the entire scene and only the subarea designated by a red circle with center $(5.5,1)\,\text{m}$ and radius $0.5\,\text{m}$.\vspace{-0.3cm}}
\label{Setup_sketch}
\end{figure}

To demonstrate the advantages of the proposed algorithm, we focus on a distributed sensing setup where the view of some radars is blocked by an obstacle, see Fig.~\ref{Setup_sketch} for an illustration. Moreover, we use the following parameter values $N=4$, $M=4$, $K=3$, $f_0=0.1$~GHz, and $B_R=150$~MHz. The capacity of the fronthaul links is modeled as $C_n = B_C \log_2\left(1+\gamma_n\right)$, where $B_C$ is the bandwidth of the fronthaul links and $\gamma_n$ is the signal-to-noise ratio of the $n$th fronthal link. We adopt $\gamma_n=\alpha_n\text{SNR}$, where $\alpha_n$ are chosen as $ 0.75,1.25,1.25,0.75$ for $n=1,\dots,4$, respectively, which enables us to vary the capacities of all fronthaul links via parameter SNR. Moreover, we choose the fronthaul bandwidth w.r.t. the sampling frequency of the radar signal as $B_C=\{1,5\}\times f_s$. 

\subsection{Performance Evaluation}

\begin{figure}[t]
\centering
\resizebox{0.74\linewidth}{!}{%
\begin{tikzpicture}[scale=1]
    \begin{semilogyaxis}[
        %style={font=\tiny},
        height=6cm,
        width=\linewidth,
        mark=none,
        xlabel={Iteration Number},
        ylabel={Average CRLB},
        grid=major,
        xmin=0, xmax=15, % Setting the range for the x-axis
        ymin=0.001, ymax=10,
        grid style=dashed, 
        %log ticks with fixed point,
        legend columns=-1,
        legend pos=south east,
        legend style={font=\scriptsize}
    ]

    \addplot[mark=none, black,forget plot] coordinates {(0,0.0058) (15,0.0058)}node[pos=0.2, below] {\scriptsize Infinite Capacity};
    \addplot[mark=none, dashed, blue,forget plot] coordinates {(0,3.7847) (15,3.7847)};
    \addplot[mark=none, blue, <->,forget plot] coordinates {(12,3.7847) (12,1.177)}node[pos=0.5,pin=180:{\tiny \color{blue}Optimisation Gain for AFVQ}] {};
    %\addplot[mark=none, dashed, red,forget plot] coordinates {(0,3.7847) (15,3.7847)};
    \addplot[mark=none, red, <->,forget plot] coordinates {(14,3.7847) (14,0.00908)}node[pos=0.7,pin=160:{\tiny \color{red} Optimisation Gain for RVQ}] {};
    \addplot[mark=none, black, <->,forget plot] coordinates {(13,0.441) (13,0.00647)} node[pos=0.8,pin=160:{\tiny \color{black}Gain of RVQ over AFVQ}] {};
    \addplot[mark=none, color={blue},forget plot] table [
        x expr=\coordindex, % Using the row index as the x value
        y index=0,
        col sep=comma,
    ] {test7/CRLB_diag_var_focused_mean.csv} node[above=-0.1,pos=0.2] [rotate=-10, font=\tiny, anchor=south west] {SNR 0 db}; 
    
    \addplot[mark=none, color={blue},forget plot] table [
        x expr=\coordindex, % Using the row index as the x value
        y index=2,
        col sep=comma,
    ] {test7/CRLB_diag_var_focused_mean.csv} node[above=-0.1,pos=0.2] [rotate=-7, font=\tiny, anchor=south west ] {SNR 10 db}; 
    
    \addplot[mark=none, color={blue}] table [
        x expr=\coordindex, % Using the row index as the x value
        y index=4,
        col sep=comma,
    ] {test7/CRLB_diag_var_focused_mean.csv} node[above=-0.1,pos=0.2] [rotate=-5, font=\tiny, anchor=south west] {SNR 20 db}; 
    
    \addplot[mark=none, color={red},forget plot] table [
        x expr=\coordindex, % Using the row index as the x value
        y index=0,
        col sep=comma,
    ] {test7/CRLB_herm_var_focused_mean.csv} node[above=-0.2,pos=0.2] [rotate=-43, font=\tiny, anchor=south west] {SNR 0 db}; 
    
    \addplot[mark=none, color={red},forget plot] table [
        x expr=\coordindex, % Using the row index as the x value
        y index=2,
        col sep=comma,
    ] {test7/CRLB_herm_var_focused_mean.csv} node[above=-0.2,pos=0.17] [rotate=-50, font=\tiny, anchor=south west] {SNR 10 db}; 
    
    \addplot[mark=none, color={red}] table [
        x expr=\coordindex, % Using the row index as the x value
        y index=4,
        col sep=comma,
    ] {test7/CRLB_herm_var_focused_mean.csv} node[above=-0.3,pos=0.15] [rotate=-55, font=\tiny, anchor=south west] {SNR 20 db}; 
    \legend{AFVQ,RVQ}
    \end{semilogyaxis}
\end{tikzpicture}
}
\caption{The achieved average CRLB (over the circle area in Fig.~\ref{Setup_sketch}) vs. the iteration number $i$ for the AFVQ and RVQ quantization schemes and different SNR values of fronthaul links, and $B_C=5f_s$. \vspace{-0.3cm}}
\label{Fig:ResCap7} 
\end{figure}
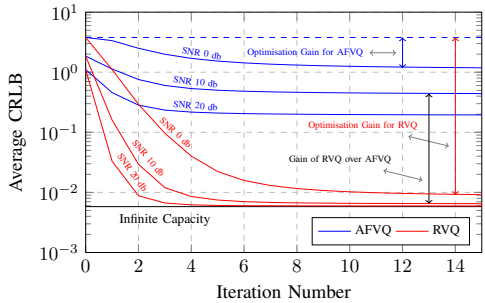

Fig.~\ref{Fig:ResCap7} shows the average CRLB (in the red circle in Fig.~\ref{Setup_sketch}) vs. the iteration number in Algorithm~\ref{alg:TDMA} for the AFVQ and RVQ quantization schemes and different SNR values of fronthaul link. As performance upper bound, we show the results for infinite-capacity fronthaul links leading to a lower bound on the achievable CRLB. To evaluate the benefits of optimizing the fronthaul compression and time allocation, we consider as a baseline scheme, the AFVQ quantization with uniform distortion and uniform power allocation, which was used to initialize Algorithm~\ref{alg:TDMA}, see Section~\ref{sec:discussion}. Fig.~\ref{Fig:ResCap7} shows that Algorithm~\ref{alg:TDMA} converges after about 5 to 15 iterations, depending on the quantization scheme and SNR of the fronthaul links. In particular, Fig.~\ref{Fig:ResCap7} suggests that the convergence is faster for higher SNRs. Moreover, for both AFVQ and RVQ quantization schemes, a significant gain is achieved compared to the benchmark by optimizing the quantization and time allocation. Furthermore, we observe from Fig.~\ref{Fig:ResCap7} that RVQ significantly outperforms AFVQ by efficiently exploiting the correlation among each radar's signals, and approaches the lower bound on CRLB when the fronthaul SNR is sufficiently large.

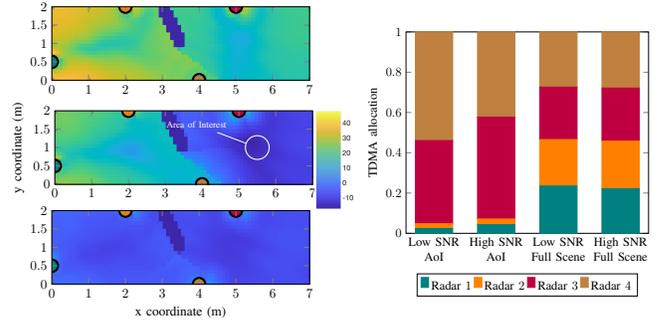
\begin{figure}[t!]
\begin{minipage}[c]{0.49\linewidth}
\begin{tikzpicture}[scale=0.5]
\begin{axis}[enlargelimits=false, axis on top, axis equal image,ylabel={\textcolor{white}{y}}]
\addplot graphics [xmin=0,xmax=7,ymin=0,ymax=2] {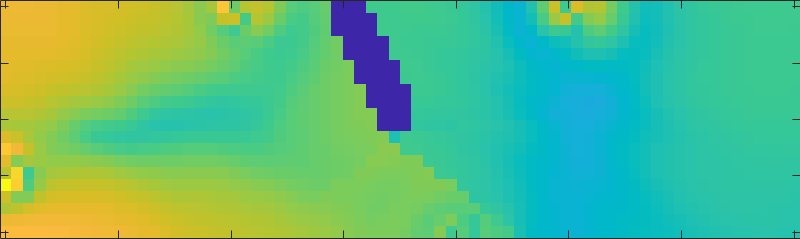};
\node at (axis cs:0,0.5) [
    circle,
    draw,
    fill,
    draw=black,fill=teal,
    ultra thick,
    minimum size=1ex
] {};
\node at (axis cs:2,2) [
    circle,
    draw,
    fill,
    draw=black,fill=orange,
    ultra thick,
    minimum size=1ex
] {};
\node at (axis cs:5,2) [
    circle,
    draw,
    fill,
    draw=black,fill=purple,
    ultra thick,
    minimum size=1ex
] {};
\node at (axis cs:4,0) [
    circle,
    draw,
    fill,
    draw=black,fill=brown,
    ultra thick,
    minimum size=1ex
] {};
\end{axis}
\end{tikzpicture}
\begin{tikzpicture}[scale=0.5]
\begin{axis}[enlargelimits=false, axis on top, axis equal image,ylabel={y coordinate (m)}]
\addplot graphics [xmin=0,xmax=7,ymin=0,ymax=2] {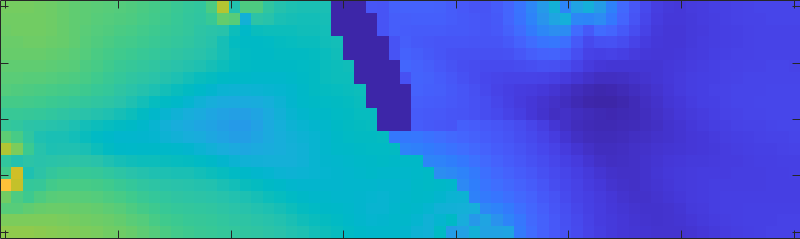};
\node at (axis cs:5.5,1) [
    circle,
    draw,
    white,
    thick,
    minimum size=4ex,
    pin={[pin edge=very thick, pin edge=white,]150:{\textcolor{white}{\scriptsize Area of Interest}}}
] {};
\node at (axis cs:0,0.5) [
    circle,
    draw,
    fill,
    draw=black,fill=teal,
    ultra thick,
    minimum size=1ex
] {};
\node at (axis cs:2,2) [
    circle,
    draw,
    fill,
    draw=black,fill=orange,
    ultra thick,
    minimum size=1ex
] {};
\node at (axis cs:5,2) [
    circle,
    draw,
    fill,
    draw=black,fill=purple,
    ultra thick,
    minimum size=1ex
] {};
\node at (axis cs:4,0) [
    circle,
    draw,
    fill,
    draw=black,fill=brown,
    ultra thick,
    minimum size=1ex
] {};
\end{axis}
\end{tikzpicture}
\begin{tikzpicture}[scale=0.5]
\begin{axis}[enlargelimits=false, axis on top, axis equal image,ylabel={\textcolor{white}{y}},xlabel={x coordinate (m)}]
\addplot graphics [xmin=0,xmax=7,ymin=0,ymax=2] {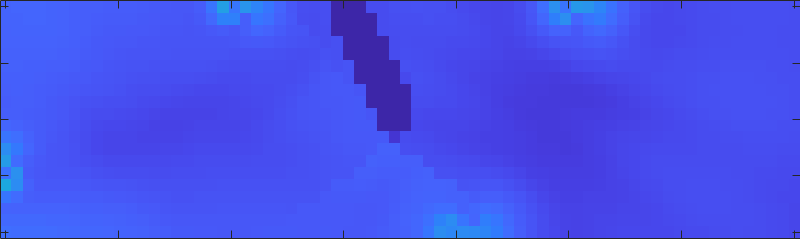};
% \node at (axis cs:5.5,1) [
%     circle,
%     draw,
%     red,
%     thick,
%     minimum size=4ex,
%     pin={[pin edge=very thick, pin edge=white,]150:{\textcolor{white}{\scriptsize Area of Interest}}}
% ] {};
\node at (axis cs:0,0.5) [
    circle,
    draw,
    fill,
    draw=black,fill=teal,
    ultra thick,
    minimum size=1ex
] {};
\node at (axis cs:2,2) [
    circle,
    draw,
    fill,
    draw=black,fill=orange,
    ultra thick,
    minimum size=1ex
] {};
\node at (axis cs:5,2) [
    circle,
    draw,
    fill,
    draw=black,fill=purple,
    ultra thick,
    minimum size=1ex
] {};
\node at (axis cs:4,0) [
    circle,
    draw,
    fill,
    draw=black,fill=brown,
    ultra thick,
    minimum size=1ex
] {};
\end{axis}
\end{tikzpicture}
\end{minipage}
\hspace{-0.5cm}
\begin{minipage}[c]{0.06\linewidth}
\includegraphics[width=\linewidth]{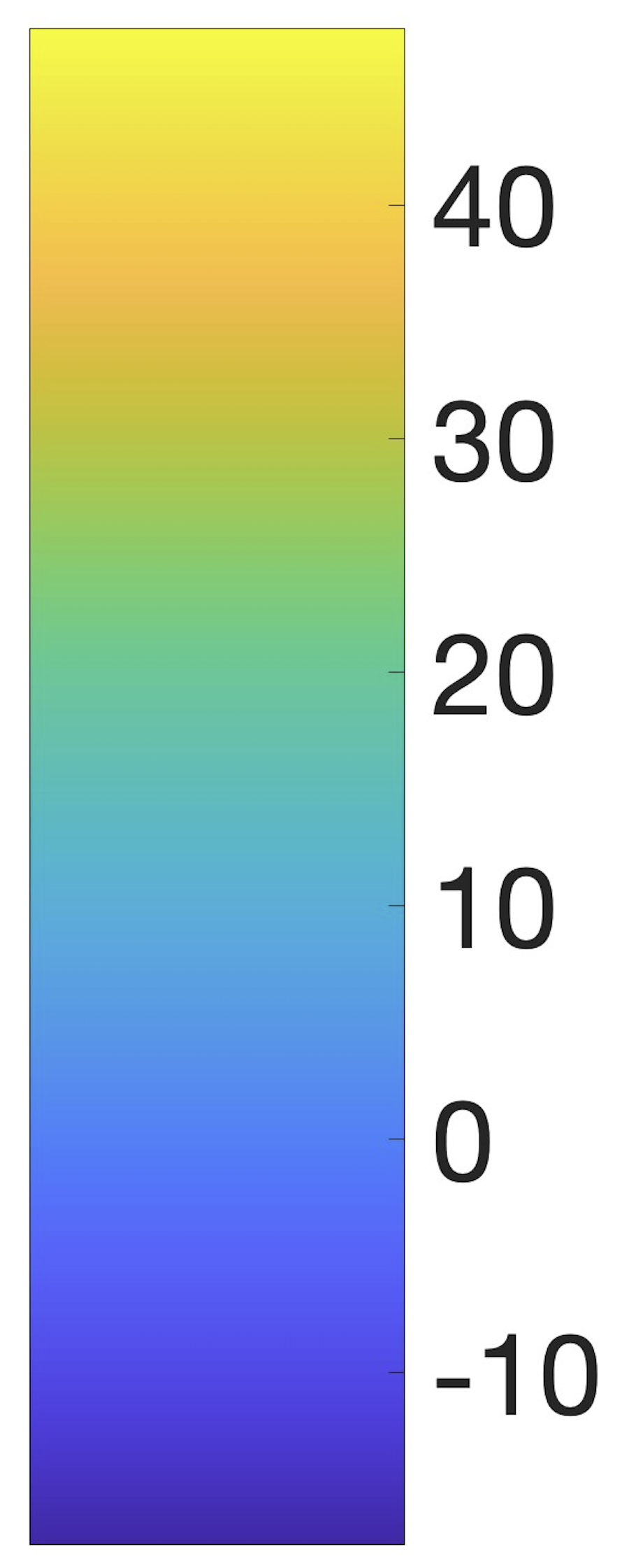}
\end{minipage}
\begin{minipage}[c]{0.45\linewidth}
\resizebox{\linewidth}{!}{%
\begin{tikzpicture}[scale=0.7]
    \begin{axis}[
        ybar stacked,
    	bar width=30pt,
        ymin=0, ymax=1,
        enlarge x limits=0.15,
        legend style={at={(0.5,-0.20)},
        anchor=north,legend columns=-1},
        ylabel={TDMA allocation},
        symbolic x coords={LFocus, HFocus, LFull, HFull},
        xtick = {LFocus, HFocus, LFull, HFull},
        xticklabels={Low SNR \\ AoI, High SNR \\AoI,Low SNR \\Full Scene,High SNR \\Full Scene},
        xticklabel style={align=center},
        ]
    \addplot+[ybar, teal] plot coordinates {(LFocus,0.0242) (HFocus,0.0431) (LFull,0.2351) (HFull,0.2220) };
    \addplot+[ybar, orange] plot coordinates {(LFocus,0.0232) (HFocus,0.028) (LFull,0.2306) (HFull,0.2357)};
    \addplot+[ybar, purple] plot coordinates {(LFocus,0.4132) (HFocus,0.5062) (LFull,0.2606) (HFull,0.2638)};
    \addplot+[ybar, brown] plot coordinates {(LFocus,0.5396) (HFocus,0.4227) (LFull,0.2736) (HFull,0.2785)};
    \legend{\strut Radar $1$, \strut Radar $2$, \strut Radar $3$, \strut Radar $4$}
    \end{axis}
\end{tikzpicture}
}
\end{minipage}
\caption{The figures on the left show the heatmap for CRLB (dB) at each location for an SNR of 0\,dB, RVQ, and $B_C = f_s$, where the top, middle, and bottom figures correspond to CRLB before optimization, after optimization w.r.t. the circle as the area of interest (AoI),  and after optimization the full scene.
The figure on the right depicts the allocation of time resources in the TDMA scheme, i.e., $a_n$, for $\textrm{SNR}=0$ dB (low SNR) and $\textrm{SNR}=30$ dB (high SNR) when optimizing for the AoI and the full scene.\vspace{-0.4cm} 
}
\label{CRLB_heatmap}
\end{figure}

The heatmaps on the left of Fig.~\ref{CRLB_heatmap} show the CRLB (dB) at each location of the scene. We see from the top figure that before optimizing the fronthaul compression and time allocation, the CRLB assumes large values for all points on the scene, particularly at some boundary points (e.g. around $(0,2)$). The middle figure shows that when we are interested in a particular area of interest (AoI), i.e., shown by the circle, Algorithm~\ref{alg:TDMA} allocates resources in such a way that the CRLB is significantly decreased for the points on the AoI, whereas other locations that are not of interest may assume large CRLB (e.g., the left side of the scene). Here, the CRLB averaged over the AoI is reduced from $13.9886$~dB before the optimization to $-14.6776$~dB after optimization. On the other hand, we see from the bottom figure that a relatively uniform CRLB is achievable when the resources are optimized for the entire scene. In this case, the  CRLB averaged over the entire scene is reduced from $27.2108$~dB before the optimization to $-6.9859$~dB after optimization. The average CRLB within the AoI in this case is $-10.4120$~dB which is much higher compared to optimizing for the AoI. Hence, prior information about the approximate target location can be exploited to significantly improve the sensing performance.

The bar charts on the left of Fig.~\ref{CRLB_heatmap} show the fraction of allocated time resources in the TDMA scheme to each radar, i.e., $a_n$. Fig.~\ref{CRLB_heatmap} reveals that when resources are allocated w.r.t. the AoI, radars 3 and 4 (see Fig.~\ref{Setup_sketch}) receive most fronthaul time due to having the best view over the AoI, whereas radars 1 and 2 receive negligible fronthaul time due to their obstructed view. Nonetheless,  the allocated time to radars 3 and 4 increases when the fronthaul capacity in higher (i.e., higher SNRs). On the other hand, when resources are optimized w.r.t. the entire scene, the times allocated to different radars become relatively uniform. The uniform allocation remains valid for both high and low SNRs since the spatial diversity of all radars is needed for the considered setup to be able to minimize the CRLB averaged over the entire scene. 
Overall, Fig.~\ref{CRLB_heatmap} underlines the importance of optimized fronthaul compression and time allocation.

\vspace{-0.1cm}
\section{Conclusions}
\vspace{-0.1cm}
In this paper, we have investigated a network of distributed radar sensors that communicate with a central processing unit over capacity-constrained fronthaul links. We have formulated a problem for optimizing the fronthaul compression and time allocation such that the CRLB is minimized for the observed scene or a specific area of interest. Moreover, as a solution to this problem, we have developed an efficient algorithm based on SDP and ACO and analyzed the convergence and complexity of the proposed algorithm. Furthermore, we presented simulation results which underline the performance gain achievable by using distributed sensing and optimized fronthaul compression under a limited fronthaul capacity.

\section*{Acknowledgment}
\vspace{-0.1cm}
This work has been performed in the context of the LOEWE center emergenCITY [LOEWE/1/12/519/03/05.001(0016)/72].

\bibliographystyle{IEEEtran} % We choose the "plain" reference style
\bibliography{ref} % Entries are in the refs.bib file

\end{document}